# Nonlinear light mixing by graphene plasmons

D. Kundys[1], B. Van Duppen[2], O. P. Marshall[1], F. Rodriguez[1], I. Torre[3,4], A. Tomadin[3], M. Polini[3*], and A. N. Grigorenko[1**]

[1]*School of Physics and Astronomy, the University of Manchester, Manchester, M13 9PL, UK*

[2]*Department of Physics, University of Antwerp, Groenenborgerlaan 171, B-2020 Antwerp, Belgium*

[3]*Istituto Italiano di Tecnologia, Graphene Labs, Via Morego 30, I-16163 Genova, Italy*

[4]*NEST, Scuola Normale Superiore, I-56126, Pisa, Italy*

*Correspondence to: Marco.Polini@iit.it

**Correspondence to: sasha@manchester.ac.uk

**Graphene is known to possess strong optical nonlinearity. Its nonlinear response can be further enhanced by graphene plasmons. Here, we report a novel nonlinear electro-absorption effect observed in nanostructured graphene due to excitation of graphene plasmons. We experimentally detect and theoretically explain enhanced nonlinear mixing of near-infrared and mid-infrared light in arrays of graphene nanoribbons. Strong compression of light by graphene plasmons implies that the effect is non-local in nature and orders of magnitude larger than the conventional local graphene nonlinearity. The effect can be used in variety of applications including nonlinear light modulators, light multiplexors, light logic, and sensing devices.**

Optical nonlinearities and inelastic light scattering observed in a continuous medium are normally *local* in nature. This means that the response of a system at a given point $r$ in space depends solely on the excitation of the system at the same point. As a result, physical parameters (e.g., dipole moment, polarization density, refractive index, etc.) at a point $r$ can be written as functions of light fields taken at the same point, provided that fields are not too large [1]. The situation can be different for a *nanostructured* medium. Strong light-matter interactions in systems with material inclusions can result in compression of light inside the inclusions [2]. In this case, optical nonlinearity or inelastic light scattering is a highly non-local and extremely large effect mediated by the excitation of internal modes of electromagnetic vibrations. One of the best-known examples of such kind is surface enhanced Raman scattering where metallic inclusions allow one to achieve unprecedented enhancement of Raman signals due to excitation of localized plasmon resonances [3,4].

Graphene is an ideal material [5] to study non-local optical non-linearity. Graphene possesses a high value of conventional optical non-linearity [6] which has been applied for pulse compression in femtosecond lasers [7]. In addition, graphene transparency can be tuned by gating voltage through the effect of optical Pauli blocking [8], thus demonstrating a strong electro-absorption effect useful for optical modulators [9]. Finally, graphene exhibits extremely large values of spatial light compression for intrinsic graphene plasmons [10-15]. As a result, graphene plasmons have been used to enhance the responsivity of mid-IR photodetectors [16], to sense surface-adsorbed polymers [17], to detect protein monolayers [18], and to modulate the emission of a terahertz quantum cascade laser [19]. Recently, there has been a great deal of theoretical and experimental interest in the interplay between plasmons and nonlinear optical properties [20] including those of graphene and its nanostructures [21-27]. An all-optical plasmon coupling scheme, which takes advantage of the intrinsic nonlinear optical response of graphene, has been implemented experimentally [28].

Here, we report non-local optical nonlinearity controlled by graphene plasmons. We predict, observe and describe a new electro-absorption effect in nanostructured graphene that generates effective nonlinear light mixing. This effect has some analogy with the quantum confined Stark effect [29] where light absorption of a quantum well can be governed by an electric field and the Franz-Keldysh effect where light absorption of a semiconductor can be induced by an applied electric field [30]. The nonlinear light mixing by graphene plasmons (NLMGP) depends on graphene doping, the geometry of structuring, temperature and could be orders of magnitude larger than that produced by the conventional local optical nonlinearity of graphene. In contrast to optical Pauli blocking (which requires the gating field to be perpendicular to the surface and the light field along the graphene surface), the light mixing discussed in our work happens for two fields that are both polarized in the plane of the nanostructured array.

Figure 1 provides the rationale behind NLMGP. A light wave that impinges on a graphene nanoribbon excites vibrations of graphene electron density, Fig. 1(a), and electron currents, Fig. 1(b). These are quite strong for a light field which is in resonance with a nanoribbon "localized" plasmon. Both of vibrations can be used to achieve light mixing: vibrations of electron density could generate light mixing through the conventional optical Pauli blocking (under a proper spatial arrangement of light beams) and vibrations of electron currents through the current-induced birefringent absorption in graphene [31]. At this stage, we concentrate our attention on NLMGP generated by currents, see Fig. 1(c). Conventional Pauli blocking also gives rise to a small light mixing signal at the frequency we are using to probe the nanostructured array, as we discuss below.

(a)

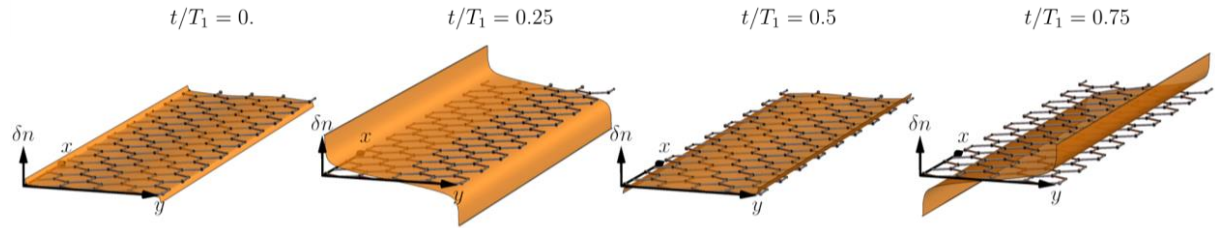

(b)

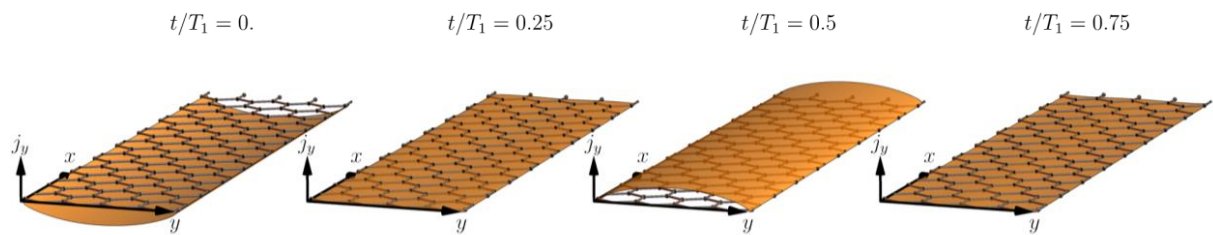

(c)

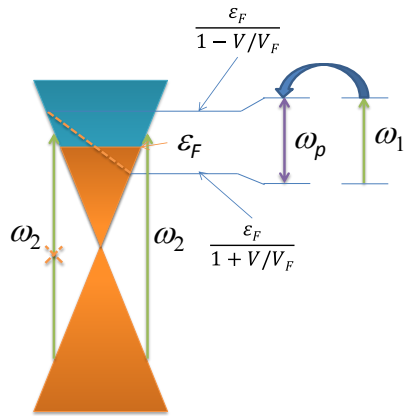

(d)

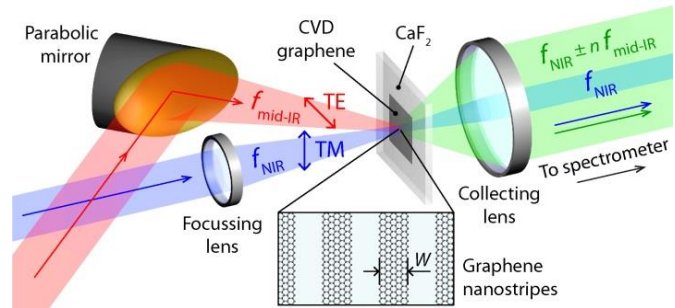

FIG. 1. Graphene plasmons and light mixing. (a) Snapshots of the temporal evolution of the electron density. (b) The current density due to the excitation of a transverse plasmon in a graphene nanoribbon. Results for different times (in units of the oscillation period $T_1$) are illustrated. Notice that the electron density profile is antisymmetric with respect to the ribbon axis. On the contrary, the velocity profile is symmetric with respect to the ribbon axis. (c) Change of graphene light absorption due to excitation of graphene plasmons. (d) Geometry of a graphene array and alignment of light beams for observation of NLPMG.

When a probe beam of angular frequency $\omega_2$ passes through a graphene nanoribbon (at zero temperature) it will be partially absorbed provided the photon energy $\omega_2$ is chosen to be larger than twice the Fermi energy, Fig. 1(c). If the graphene nanoribbon is additionally illuminated by a pump beam of angular frequency $\omega_1$ (chosen to be close to the plasmon frequency $\omega_{\text{pl}}$) the currents in graphene tilt the Fermi level (the dashed yellow line) and forbid some transitions due to the optical Pauli blocking, see Fig. 1(c). The spectral region in which transitions can be blocked/unblocked is defined by $\frac{2 E_F/\hbar}{1+v_y/v_F} < \omega < \frac{2 E_F/\hbar}{1-v_y/v_F}$, where $v_y$ is the drift velocity proportional to the induced current and $E_F$ is the Fermi level. Since the current, and hence the drift velocity, changes periodically with the pump field, Fig. 1(b), this implies that the absorption of the probe beam by the graphene nanoribbon will be modulated with the pump beam frequency resulting in effective nonlinear and non-local light mixing. The maximal amplitude of the modulation achieved with this mechanism is $\pi\alpha/2$, where $\alpha \approx 1/137$ is the quantum electrodynamics fine structure constant, and depends on the incident light polarizations and frequencies, the geometry of structuring, the chemical potential of graphene and the electron temperature (see Supplementary Information). The main advantage of NLMGP is that the upper boundary of the spectral range can become very large if $v_y$ approaches $v_F$ such that it is possible, at least in principle, to modulate lasers with large photon energy by using a system with relatively low electron concentration. Figure 1(d) shows the basic elements of NLMGP used in our experiments.

To demonstrate practically useful NLMGP, we have chosen a telecom probe beam and a mid-infrared pump beam. The schematics of the setup is shown in Fig. 2. A stabilized $CO_2$ laser with a CW output power of maximum 200 mW and wavelength of 10.6 μm was used to excite plasmons in graphene nanoribbons (see below). A wavelength-tunable 1520-1630 nm Agilent 81949A compact C+L band telecom laser was used to produce the probe beam. To image the sample, we have employed an *ad hoc* microscope consisting of an 8% reflectivity pellicle beam splitter, a high-magnification 12X zoom lens with a 12 mm fine focus system and a CMOS camera; the white light from the fiber illuminator was launched along the main optical axis with a 30% reflectivity beam splitter for simultaneous sample imaging and beam alignment. Both beam splitters were flip-mounted to be used only during the alignment and were removed during the wave-mixing measurements.

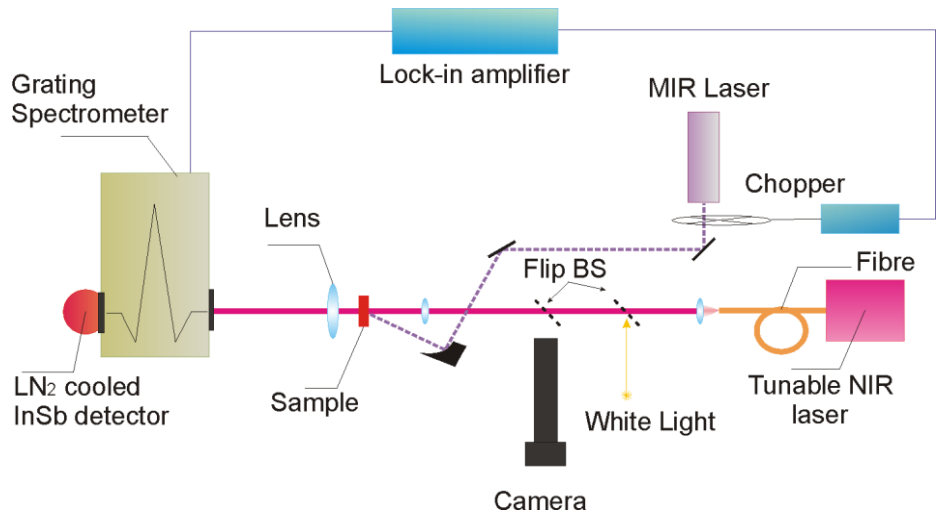

FIG. 2. Schematic diagram of the experimental setup used for measuring NLMPG.

The two laser beams were focused on a 50 μm spot using a low numerical aperture (NA) lens and a parabolic mirror (with NA = 0.12 and 0.05, respectively), were overlapped in space with a pinhole and then repositioned over the sample. We estimate the electric field provided by the $CO_2$ laser to be ~3 kV/cm. The angle between the two lasers incident beams was 23°. In this geometry the resulting light mixing $|k_{NIR} + nk_{mid\ IR}|$ signals will propagate at small angles of < 3° and < 6° to normal incidence for the first- and second-order sidebands, respectively. In our experiment we choose to discriminate sidebands spectrally and, therefore, have used a high numerical aperture collection lens ($NA = 0.5$) in order to ensure efficient light collection at angles of up to 26.5°. The collected light has been analysed by a Cornerstone 130 monochromator with a liquid-nitrogen-cooled amplified InSb infrared detector and measured using a low noise lock-in detection technique.

NLMPG was studied in arrays of nanoribbons made out of graphene grown by chemical vapor deposition (CVD) with the stripe width $w$ ranging from 20 nm to 50 nm and a nominal 50/50 inter-ribbon duty cycle, see Supplementary Information. The set of samples with different stripe widths $w$ were fabricated by electron beam lithography and their optical properties were measured. The sample arrays that exhibited the localized graphene plasmon resonance at the wavelength of the pump laser were chosen for graphene light mixing.

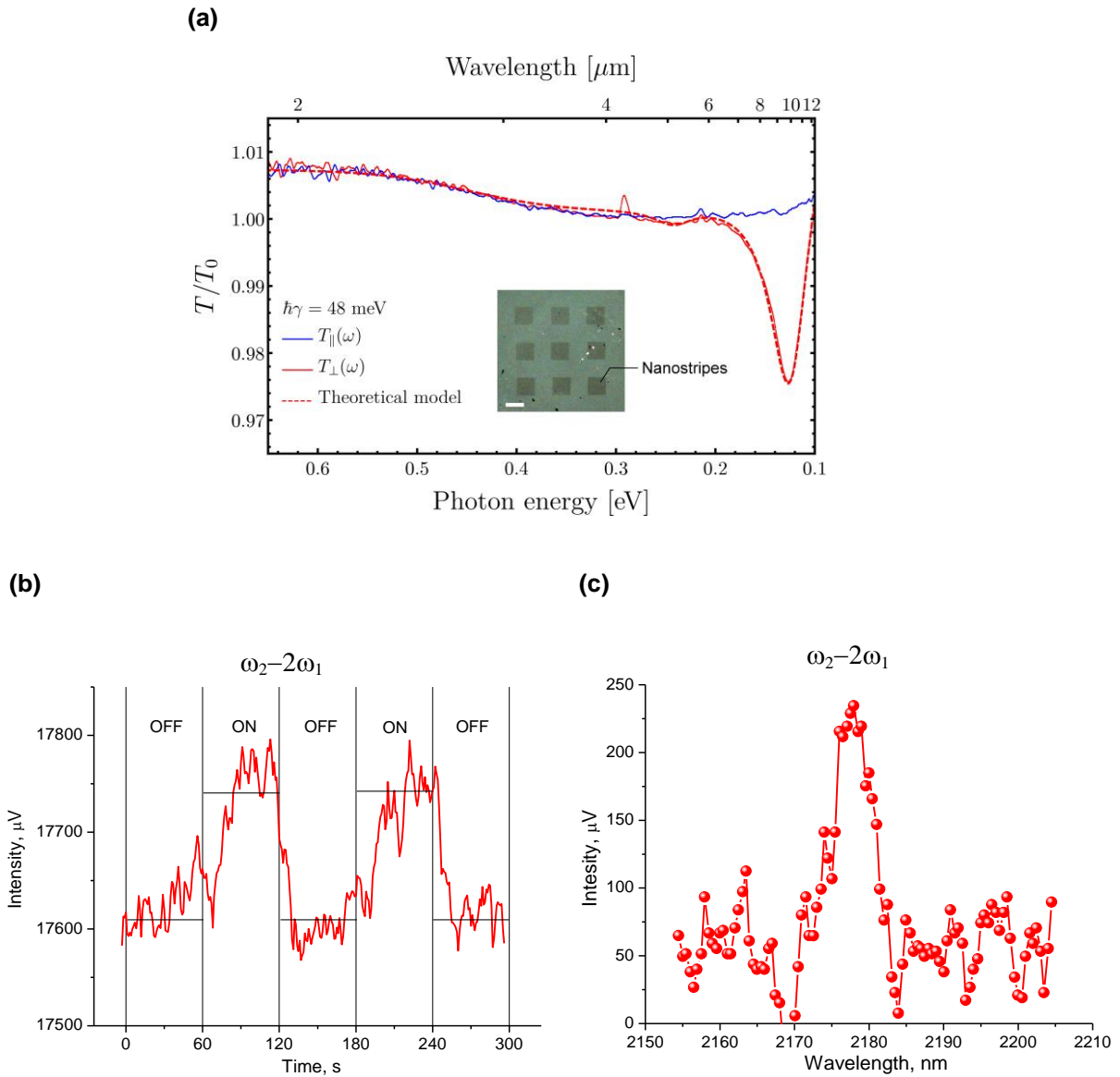

FIG. 3. Light mixing with graphene plasmons. (a) Transmission spectra of the array ($T$), relative to the neighboring unpatterned graphene ($T_0$), measured at normal incidence for two different polarizations. The transverse plasmon absorption is visible at ~126 meV for TE polarized light. The small artifact around 290 meV is due to atmospheric $CO_2$ absorption. The fitting of the plasmon resonance is shown as the dashed curve. The fitting parameters are $\hbar\gamma$ = 48 meV, $w$ = 50 nm, $\epsilon$ = 1.8, $E_F$ = 0.22 eV.

Inset shows an optical microscope image of several nanoribbon arrays. Scale bar = 50 μm. (b) A mixing signal at the combination frequency $\omega_2 - 2\omega_1$ for the probe (wavelength 1.5μm) and pump (10.6μm) beam powers of 4 mW and 200 mW. (c) The spectral dependence near the sideband $\omega_2 - 2\omega_1$.

Figure 3(a) shows the representative TE and TM transmission spectra measured on a nanoribbon array which is resonant with the $CO_2$ laser used in the experiments. The spectra were acquired using a Brucker FTIR spectrometer and microscope. (To avoid ambiguity, TE transmission with electric field of incident light perpendicular to the stripes is labelled $T_\perp(\omega)$; while TM transmission with electric field parallel to the stripes is labelled $T_\parallel(\omega)$.) The drop in $T_\perp(\omega)$ light transmission at wavelengths of ~10 μm corresponds to the plasmon resonance of the nanoribbons. The dashed curve in Fig. 3(a) presents an excellent fit for $T_\perp(\omega)$ calculated with the theory described in detail in Supplementary Information. Using our knowledge of the geometric width ($w = 50$ nm) of the measured ribbons, we find an effective dielectric constant $\epsilon = 1.8$ and a phenomenological plasmon broadening $\hbar\gamma = 48$ meV from the fit, see below. The chemical potential of the graphene was evaluated to be $E_F \sim 0.2$ eV from fitting the optical absorption step in Fig. 3(a). In such fitting we took into account the reduced fractional areal coverage ($A_{fill}$) of graphene in the nanoribbon array compared to the unpatterned background. This step height is marginally smaller than the $\pi\alpha/2$ value expected for the 50/50 inter-ribbon duty cycle, suggesting $A_{fill} \sim 0.7$, most likely due to unintentional inclusion of some unpatterned graphene in the transmission measurement area.

Figures 3(b) and (c) show typical NLMGP results. Figure 3(b) plots the signal at the combination frequency $\omega_2 - 2\omega_1$, when the pump laser is either on or off. The light mixing signal was measured using the probe and pump beam powers of 4 mW and 200 mW, respectively. The plasmon-assisted nonlinear conversion coefficient $\eta$ corresponding to the measured signal is of the order $\eta = 1.8 \times 10^{-5}$ which is two orders of magnitude larger than the corresponding quantity measured in a continuous graphene sheet [6]. Based on the data from the experiments, we calculate an effective third-order nonlinearity $\chi^{(3)} = 4.5 \times 10^{-6}$ (esu) which is one order of magnitude larger than the corresponding quantity measured in a continuous graphene sheet [6]. The spectral dependence near the sideband $\omega_2 - 2\omega_1$ is shown in Fig. 3(c). A peak is clearly visible at the combinational wavelengths, thus eliminating possible thermal effects as a source of the signal in Fig. 3(b).

Now we describe our theoretical approach to NLMGP. A graphene nanoribbon supports "running" plasmons that propagate along the longitudinal ribbon direction and "localized" transverse plasmons in which the electron liquid oscillates back and forth between the ribbon edges, transversally to the ribbon axis. Here, we are interested in the latter type of plasmons, which, for the sake of simplicity, will be referred to as "transverse" plasmons. In the realm of linear response theory [32], the transverse plasmon mode with the *lowest* energy can be described by the following relation

$$\hbar\omega_{\mathrm{pl}} = \sqrt{\frac{2\,\xi\,\alpha_{\mathrm{ee}}\,\hbar v_{\mathrm{F}}}{\epsilon\,w}} E_{\mathrm{F}}, \qquad (1)$$

where $\hbar\omega_{\mathrm{pl}}$ is the plasmon energy, $\xi = 2.31$ is a numerical constant, $\alpha_{\mathrm{ee}} = 2.2$ is the so-called graphene fine structure constant [32], $v_{\mathrm{F}}$ is the graphene Fermi velocity, $E_{\mathrm{F}}$ is the Fermi energy, $w$ is the width of the nanoribbon, and $\epsilon$ is an effective dielectric constant. Macroscopically, transverse plasmons are described by space- and time-dependent current density $j_y(\mathbf{r}, t)$ and carrier density $\delta n(\mathbf{r}, t)$ profiles, which can be calculated by treating the pump laser oscillating at frequency $\omega_1$ in the framework of linear response theory (see Supplementary Information). These quantities along with the directions of *x*- and *y*-axes are shown in Figs. 1(a) and (b) at different times during one cycle of the pumping field. The resonant response of electron plasma in graphene happens when $\omega_1$ matches $\omega_{\text{pl}}$.

Our target is to analyze the response of the ribbon array in the presence of both a pump laser at frequency $\omega_1$ and a probe laser at frequency $\omega_2$. The mixing of the two laser lights generates a NLMGP signal at the frequency combination $\omega_2 - 2\omega_1$ (among others) which can be explained in terms of a third-order nonlinear optical process enabled by the graphene nanostructure. The signal measured by the spectrometer is related to the third-order polarization $\boldsymbol{P}^{(3)}(\omega_2 - 2\omega_1)$, whose $x$ component is given by [1]

$$P_x^{(3)}(\omega_2 - 2\omega_1) =$$
$$\frac{D}{4} \int d\boldsymbol{r} \int d\boldsymbol{r}' \int d\boldsymbol{r}'' \chi_{xyyx}^{(3)}(\boldsymbol{r},\boldsymbol{r}',\boldsymbol{r}'';\omega_2 - 2\omega_1) E_{1,y}^*(\boldsymbol{r}'',\omega_1) E_{1,y}^*(\boldsymbol{r}',\omega_1) E_{2,x}(\boldsymbol{r},\omega_2). \quad (2)$$

In Eq. (2), $D = 3$ is the number of distinct permutations that generate the sideband at $\omega_2 - 2\omega_1$, and $E_{1,y}(\boldsymbol{r}, \omega_1)$ and $E_{2,x}(\boldsymbol{r}, \omega_2)$ are the complex amplitudes of the electric fields of the pump and probe laser, respectively. Notice that, in the experiment, the pump laser is polarized in the $y$ direction and the probe laser is polarized in the $x$ direction. Therefore, we are probing only the $\chi_{xyyx}^{(3)}$ element of the third-order susceptibility tensor (which is typically larger than $\chi_{yyyy}^{(3)}$). We emphasize the highly *non-local* nature of Eq. (2): the NLMGP signal depends on the spatial distribution of the pump field over the *whole* sample. The third-order susceptibility which is probed in our experiment is *not* the local quantity of the form $\chi_{xyyx}^{(3)}(\boldsymbol{r}, \boldsymbol{r}', \boldsymbol{r}''; \omega_2 - 2\omega_1) \propto \delta(\boldsymbol{r} - \boldsymbol{r}')\delta(\boldsymbol{r} - \boldsymbol{r}'')$. The polarization expressed in Eq. (2) is the average polarization of a medium constructed by assigning an effective thickness $d_{\text{gr}} = 0.33$ nm to the graphene nanoribbons. Since graphene is a two-dimensional material, it is, however, more natural to calculate a current density $\boldsymbol{J}(\omega_2 - 2\omega_1)$ in the graphene sheet.

The relation between the current density and the corresponding effective polarization then follows from the relation $\boldsymbol{P}(\omega) = i\boldsymbol{J}(\omega)/(\omega d_{\text{gr}})$.

To calculate graphene plasmon light mixing we use the following approach. We notice that the mixing signals arise in our system when the probe light interacts with nanoribbons whose local carrier density and current density are time-varying due to the excitation of transverse plasmons by the pump light. Since $\omega_2 \gg \omega_1$, we can separate time scales and consider the graphene electron system to be in a quasi-stationary regime from the point of view of the probe laser. This allows us to calculate the current density $J_x(\boldsymbol{r}, \omega_2)$ generated by the probe laser at frequency $\omega_2$ as

$$J_x(\boldsymbol{r}, \omega_2) = \sigma_{xx}(\boldsymbol{r}, \omega_2) E_{2,x}(\boldsymbol{r}, \omega_2), \tag{3}$$

where $\sigma_{xx}(\boldsymbol{r}, \omega_2)$ is the longitudinal optical conductivity of graphene. It is necessary to stress that this conductivity of graphene is a function of the local electron density $n(\boldsymbol{r}, t)$ and current density $\boldsymbol{j}(\boldsymbol{r}, t)$ and the separation of time scales allows us to treat time in Eq. (3) as a parameter; i.e., a label attached to $n(\boldsymbol{r}, t)$ and $\boldsymbol{j}(\boldsymbol{r}, t)$. NLMGP occurs because $n(\boldsymbol{r}, t)$ and $\boldsymbol{j}(\boldsymbol{r}, t)$ depend non-linearly and non-locally on the pump field and change periodically with frequency $\omega_1$. We then expand the conductivity $\sigma_{xx}(\boldsymbol{r}, \omega_2)$ up to *second* order in the current and carrier densities as

$$\sigma_{xx}(\boldsymbol{r}, \omega_2) \approx \sigma_{xx}^0(\omega_2) + \sigma_{xx}^{n,1}(\omega_2) \frac{\delta n(\boldsymbol{r}, t)}{\bar{n}} + \sigma_{xx}^{n,2}(\omega_2) \left(\frac{\delta n(\boldsymbol{r}, t)}{\bar{n}}\right)^2$$
$$+ \sigma_{xx}^{j,2}(\omega_2) \left(\frac{j_y(\boldsymbol{r}, t)}{\bar{j}}\right)^2. \tag{4}$$

Here $\bar{n}$ is the uniform carrier density, $\bar{J} = -ev_\text{F}\bar{n}$ is the critical current and $\sigma_{xx}^0$ is the graphene conductivity in the absence of plasmons. The coefficients $\sigma_{xx}^{n,1}, \sigma_{xx}^{n,2}$ and $\sigma_{xx}^{j,2}$ depend in general on $\bar{n}$, the probe frequency $\omega_2$, the electron temperature $T$, and the angle between the probe field and the nanoribbon [31].

Equation (4) provides the basis for calculations of NLMGP as the conductivity at the probe light frequency additionally oscillates with the frequency components $m\,\omega_1$, where $m$ is an integer. Substitution of Eq. (4) into Eq. (3) shows that the oscillation of the conductivity at frequency $\omega_1$ mixes with the oscillation at frequency $\omega_2$ induced by the probe laser. Therefore, the total local current density has Fourier components at frequencies $\omega_2 \pm m\,\omega_1$. Averaging over the graphene nanoribbon array can suppress some of the harmonics due to symmetry. The NLMPG signal shown in Fig. 3 corresponds to $m = 2$. This response is generated by the last two terms in Eq. (4), which are quadratic in the pump field $\boldsymbol{E}_1$. From these two terms, we can separate out the pump field dependence and identify the third-order conductivity $\sigma^{(3)}$ that relates the current density $\boldsymbol{J}^{(3)}(\omega_2 - 2\omega_1)$ to the product of the probe field $\boldsymbol{E}_2$ and the square of the pump field $\boldsymbol{E}_1$ as explained in the Supplementary Information. The third-order susceptibility finally follows as

$$\chi_{xyyx}^{(3)}(\boldsymbol{r},\boldsymbol{r}',\boldsymbol{r}'';\omega_2 - 2\omega_1) = \frac{i4\sigma_{xyyx}^{(3)}(\boldsymbol{r},\boldsymbol{r}',\boldsymbol{r}'';\omega_2 - 2\omega_1)}{D(\omega_2 - 2\omega_1)d_\text{gr}}. \qquad (5)$$

This analysis shows that the third-order response depends on the current density and carrier density profiles induced by transverse plasmons. We stress again that since plasmons are collective excitations of the electron liquid, these profiles depend on the response of the system as a whole and are, therefore, intrinsically non-local. We have checked that sidebands at $\omega_2 \pm 2\omega_1$ emerging from the term in Eq. (4) which is quadratic in $\delta n(\boldsymbol{r}, t)$ (corresponding to optical Pauli blocking) are parametrically small compared to those stemming from the last term in Eq. (4) (corresponding to current-induced birefringent absorption). Finally, we note that Eq. (4) also contains a term that is linear in $\delta n(\mathbf{r},t)$. Following the same logic as above, this could introduce a second-order non-local signal at frequencies $\omega_2 \pm \omega_1$. However, as the spectrometer in the experiment measures the spatial average over the entire sample, this term vanishes due to the oddness of the density profile $\delta n(\mathbf{r},t)$ with respect to the center of the nanoribbon, as shown in Fig. 1(a). Conversely, if inversion symmetry is broken (for example by fabrication or disorder), this term does not vanish and sidebands at $\omega_2 \pm \omega_1$ are expected, see Supplementary Information.

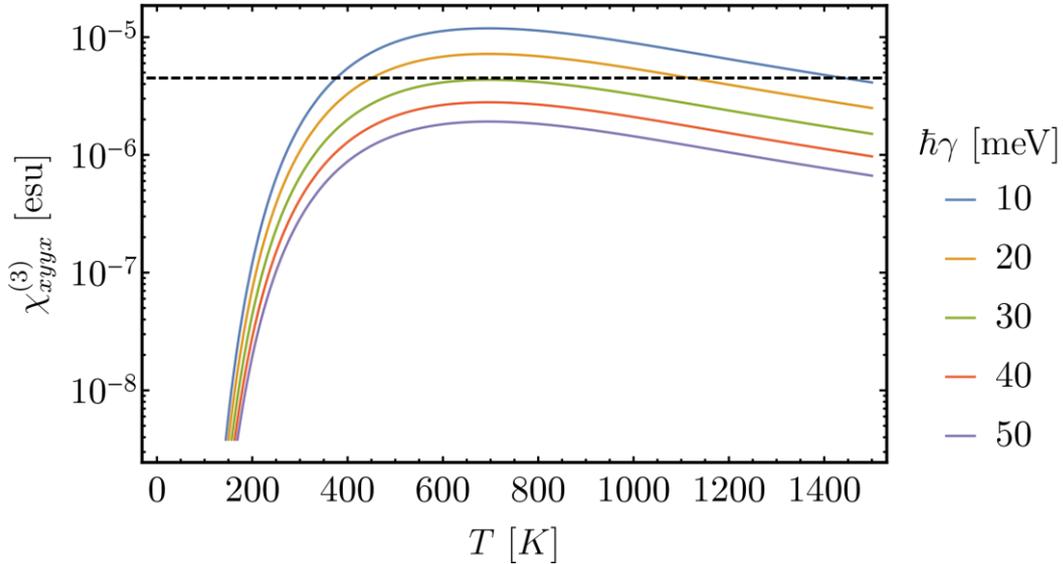

FIG. 4. Dependence of the effective third-order susceptibility on the temperature $T$ and damping $\gamma$. The other parameters of the system are as inferred from the fit of the data in Fig. 3(a). The experimentally measured value for effective $\chi^{(3)}$ is shown as a horizontal dashed line. In calculations we assumed that the inter-ribbon distance equals its width.

Figure 4 shows the calculated effective third-order susceptibility of a graphene nanoribbon array with parameters extracted from fitting the transmission data in Fig. 3(a). The effective third-order susceptibility is shown as a function of electron temperature for different values of the phenomenological damping parameter $\gamma$. The horizontal dashed line is the measured value of the third-order susceptibility. We see that our theory explains the amplitude of the measured effect very well, provided we assume that the electron gas is heated above room temperature ($T \sim 600$ K) and employ a damping parameter on the order of what is inferred from the fitting of $T_\perp(\omega)$ discussed above, see Fig. 3(a), i.e. $\hbar\gamma \sim 30$ meV-50 meV. Crude measurements of graphene temperature described in Supplementary Information show that the graphene array reaches temperatures well above 500 K in our experiments.

To conclude, we suggest, observe and describe a new nonlinear electro-absorption effect in nanostructured graphene which is orders of magnitude larger than the conventional graphene nonlinearity. Our work provides the proof-of-principle experiment in which graphene plasmons strongly affect nonlinear properties of graphene metasurface. Our results stress the potential of graphene plasmonics for development of light modulators, light multiplexors, light logic that can be used in optical interconnects, and highlights the unexpected optical phenomena possible in a system with a collective motion of electron plasma in a two-dimensional plane.

**Acknowledgments:** This work was supported by the European Union's Horizon 2020 research and innovation programme under grant agreement No.~696656 "GrapheneCore1", Bluestone Global Technology, and Fondazione Istituto Italiano di Tecnologia. BVD is supported by a postdoctoral fellowship granted by FWO-Vl.

**References:**

[1] R. W. Boyd, *Nonlinear optics* (Academic press, 2003).
[2] S. A. Maier, *Plasmonics: Fundamentals and Applications* (Springer, 2007).
[3] M. Fleischmann, P. J. Hendra, and A. McQuillan, Chemical Physics Letters **26**, 163 (1974).
[4] S. Nie and S. R. Emory, Science **275**, 1102 (1997).
[5] A. K. Geim and K. S. Novoselov, Nat Mater **6**, 183 (2007).
[6] E. Hendry, P. J. Hale, J. Moger, A. Savchenko, and S. Mikhailov, Physical Review Letters **105**, 097401 (2010).
[7] F. Bonaccorso, Z. Sun, T. Hasan, and A. C. Ferrari, Nat. Photonics **4**, 611 (2010).
[8] F. Wang, Y. B. Zhang, C. S. Tian, C. Girit, A. Zettl, M. Crommie, and Y. R. Shen, Science **320**, 206 (2008).
[9] M. Liu, X. Yin, E. Ulin-Avila, B. Geng, T. Zentgraf, L. Ju, F. Wang, and X. Zhang, Nature **474**, 64 (2011).
[10] M. Jablan, H. Buljan, and M. Soljačić, Physical Review B **80**, 245435 (2009).
[11] J. Chen *et al.*, Nature **487**, 77 (2012).
[12] Z. Fei *et al.*, Nature **487**, 82 (2012).
[13] F. H. Koppens, D. E. Chang, and F. J. Garcia de Abajo, Nano Letters **11**, 3370 (2011).
[14] A. Grigorenko, M. Polini, and K. Novoselov, Nat. Photonics **6**, 749 (2012).
[15] T. Low and P. Avouris, ACS Nano **8**, 1086 (2014).
[16] M. Freitag, T. Low, W. Zhu, H. Yan, F. Xia, and P. Avouris, Nature communications **4** (2013).
[17] Y. Li, H. Yan, D. B. Farmer, X. Meng, W. Zhu, R. M. Osgood, T. F. Heinz, and P. Avouris, Nano Letters **14**, 1573 (2014).
[18] D. Rodrigo, O. Limaj, D. Janner, D. Etezadi, F. J. G. de Abajo, V. Pruneri, and H. Altug, Science **349**, 165 (2015).
[19] S. Chakraborty, O. Marshall, T. Folland, Y.-J. Kim, A. Grigorenko, and K. Novoselov, Science **351**, 246 (2016).
[20] M. Kauranen and A. V. Zayats, Nat. Photonics **6**, 737 (2012).
[21] S. Mikhailov, PHYSICAL REVIEW B **84**, 045432 (2011).
[22] S. Mikhailov, Physical Review Letters **113**, 027405 (2014).
[23] X. Yao, M. Tokman, and A. Belyanin, Physical Review Letters **112**, 055501 (2014).
[24] M. T. Manzoni, I. Silveiro, F. G. de Abajo, and D. E. Chang, New J Phys **17**, 083031 (2015).
[25] M. Tokman, Y. Wang, I. Oladyshkin, A. R. Kutayiah, and A. Belyanin, Physical Review B **93**, 235422 (2016).
[26] J. D. Cox, I. Silveiro, and F. J. García de Abajo, ACS Nano **10**, 1995 (2016).
[27] H. Rostami, M. I. Katsnelson, and M. Polini, Physical Review B **95**, 035416 (2017).
[28] T. J. Constant, S. M. Hornett, D. E. Chang, and E. Hendry, Nature Physics **12**, 124 (2016).
[29] D. Miller, D. Chemla, T. Damen, A. Gossard, W. Wiegmann, T. Wood, and C. Burrus, Physical Review Letters **53**, 2173 (1984).
[30] D. Miller, D. Chemla, and S. Schmitt-Rink, Physical Review B **33**, 6976 (1986).
[31] B. V. Duppen, A. Tomadin, A. N. Grigorenko, and M. Polini, 2D Materials **3**, 015011 (2016).
[32] G. F. Giuliani and G. Vignale, *Quantum Theory of the Electron Liquid* (Cambridge University Press, Cambridge, 2005).